\documentclass[lettersize,journal]{IEEEtran}
\usepackage{graphicx}
\ifCLASSOPTIONcompsoc
    \usepackage[caption=false, font=normalsize, labelfont=sf, textfont=sf]{subfig}
\else
\usepackage[caption=false, font=footnotesize]{subfig}
\fi
\usepackage[referable]{threeparttablex}
\usepackage{multirow}
\usepackage{adjustbox}
\usepackage[table,xcdraw,dvipsnames]{xcolor}
\usepackage{longtable}
\usepackage{color}
\usepackage{xcolor,colortbl}
\usepackage{tikz}

\usepackage{pifont}
\usepackage{comment}
\usepackage[activate]{microtype}
\usepackage[colorlinks=false]{hyperref}

\usepackage{authblk}
\usepackage{booktabs}

\usepackage{url}
\usepackage[dvipsnames]{xcolor}
\usepackage{soul}

\usepackage[separate-uncertainty = true,multi-part-units = repeat]{siunitx}
\usepackage{array,amsfonts,amsmath}
\usepackage{caption}
\usepackage{verbatim}
\ifCLASSOPTIONcompsoc
 \usepackage[nocompress]{cite}
\else
 \usepackage{cite}
\fi
\ifCLASSINFOpdf
\else
\fi

\begin{document}


\title{Lightweight Toxicity Detection in Spoken Language: A Transformer-based Approach for Edge Devices}

\author{Ahlam Husni Abu Nada, Siddique Latif, Junaid Qadir
\thanks{Ahlam Husni Abu Nada and Junaid Qadir are affiliated with Qatar University, Doha.}
\thanks{Siddique Latif is affiliated with Queensland University of Technology (QUT), Australia.}\thanks{Email: siddique.latif@qut.edu.au}}





\maketitle

\begin{abstract}

Toxicity is a prevalent social behavior that involves the use of hate speech, offensive language, bullying, and abusive speech. While text-based approaches for toxicity detection are common, there is limited research on processing speech signals in the physical world. Detecting toxicity in the physical world is challenging due to the difficulty of integrating AI-capable computers into the environment. We propose a lightweight transformer model based on wav2vec2.0 and optimize it using techniques such as quantization and knowledge distillation. Our model uses multitask learning and achieves an average macro F1-score of 90.3\% and a weighted accuracy of 88\%, outperforming state-of-the-art methods on DeToxy-B and a public dataset. Our results show that quantization reduces the model size by almost 4 times and RAM usage by 3.3\%, with only a 1\% F1 score decrease. Knowledge distillation reduces the model size by 3.7 times, RAM usage by 1.9, and inference time by 2 times, but decreases accuracy by 8\%. Combining both techniques reduces the model size by 14.6 times and RAM usage by around 4.3 times, with a two-fold inference time improvement. Our compact model is the first end-to-end speech-based toxicity detection model based on a lightweight transformer model suitable for deployment in physical spaces. The results show its feasibility for toxicity detection on edge devices in real-world environments.

\end{abstract}



\section{Introduction} \label{sec:introduction}

Toxic behavior is a social phenomenon that occurs in places where people gather, both virtually and physically. This can include social platforms and online gaming communities, as well as schools, offices, and homes \cite{toxicinherently} \cite{toxicgaming}. Such toxicity can target different groups and individuals, especially minority communities and individuals \cite{toxicworkplaces}. In addition to having a negative psychological impact on individuals, toxic language is used in several scenarios when making threats or being violent \cite{toxicimplications}; some laws classify such language as a crime because it promotes discrimination and violence. This motivates the researchers to look for effective ways to detect toxicity to govern language and maintain a healthy environment.

Traditional methods of identifying and managing toxic language often require human oversight, which can be both labor-intensive and time-consuming, making it difficult to scale effectively. While a formal and precise definition of ``toxic language" is lacking, it generally refers to any expression that is abusive, discriminatory, hateful, offensive, or racist. Artificial intelligence (AI)-based automated systems have emerged as a promising solution for identifying and addressing toxic language in both physical and virtual spaces. 

As the Internet continues to grow exponentially, social networking platforms are expanding rapidly, resulting often in toxic user interactions. Several studies have been conducted to detect hate content on social media platforms such as Twitter, Facebook, and online gaming, including cyberbullying \cite{cyberbullying}, hate speech \cite{hatespeech}, offensive speech \cite{offensive}, and sentiment detection \cite{sentiment}. All of which can be classified as a single form of toxic language that focuses on a single feature of toxicity. Detecting toxic speech in its general meaning has also been the subject of studies that have suggested detection techniques based on various machine learning (ML) models \cite{toxiccomment1, toxiccomment2}. 

Existing research on combating toxic language has, for the most part, focused on analyzing textual content such as comments and posts using natural language processing (NLP). Recent advances have prompted increased interest in strategies that focus on multimedia audiovisual data. In this paper, we focus on sound, which is a rich source of information that can provide valuable insights into the speaker and their surroundings \cite{soundRichSource}. Despite the potential benefits of using audio data to detect toxic language, existing research in both online and physical settings has been limited. This is particularly true in physical spaces, as it has traditionally been very challenging to seamlessly integrate computers capable of running AI and ML models into the physical environment---however, this has been changing recently as embedded ML and edge AI become common \cite{situnayake2022ai,warden2019tinyml}.




Developing models for detecting toxic audio presents a significant challenge for researchers.
The use of traditional simple techniques has resulted in low performance in detecting toxicity, as evidenced by previous studies \cite{baseline1audio}. To enable accurate decision-making while accounting for diverse factors such as acoustic and linguistic information (including semantics, emotion, and tone), it is necessary to develop robust representations of speech signals. These representations must be invariant to speaker variability, background noise, reverberation, and other natural speech variations \cite{latif2021survey}. To achieve this, a comprehensive representation of the acoustic context of an utterance is required, which necessitates capturing both global and local features. In recent times, transformer-based models have proven to be highly effective in modeling long-range dependencies in speech and have surpassed traditional models such as convolutional neural networks (CNNs) and recurrent neural networks (RNNs) in a variety of speech-related tasks, including speech recognition, emotion recognition, and speaker identification \cite{transformerwhy, transformerwhy2,latif2023transformers}. 

Furthermore, detecting toxic language in real-world contexts through audio modality presents an additional challenge. Specifically, the deployment of the trained model on an edge device with limited resources is required to overcome the limitations associated with using the traditional method of manipulating data through the cloud. Thus, the proposed model must adhere to these constraints without significantly sacrificing performance.

To address the aforementioned challenges, we have proposed in this paper a lightweight transformer-based model that captures the contextualized features
that are crucial for accurately recognizing toxicity signs from speech. We use the multi-task learning (MTL) technique to enhance the model's generalization capabilities without increasing the system’s complexity, enabling it to adapt better to new and diverse data.
\subsection{Contributions of this Paper}

In this paper, we propose to the best of our knowledge the first end-to-end speech-based toxicity detection model that is trained using multi-task learning (MTL). We outline the major contributions of this work next. 
\begin{enumerate}
\item In contrast to the previous studies, we propose a multitasking framework combining ASR to construct an auxiliary task of toxicity classification. We used the pre-trained wav2vec2.0 model to make use of self-supervised learning. 
\item Evaluate the proposed model on the DeToxy-B and IEMOCAP datasets after adjusting the IEMOCAP labels to fit the toxicity problem, resulting in state-of-the-art toxicity detection results on both datasets.
\item We apply a number of model compression techniques including quantization, knowledge distillation, and pruning to make the proposed model efficient enough to be deployed on an edge device. 
\item An analysis of the compressed model's accuracy, latency, hard disk size, and memory usage on a CPU laptop, showing that the model is competitive in edge computing applications, enabling their deployment in physical settings.
\end{enumerate}

\subsection{Organization of this Paper}

The rest of the paper is structured as follows. Section \ref{sec:background} covers the underlying
concepts and background information on which the suggested solution is built and  explores the literature and cutting-edge solutions. Section \ref{sec:Proposed-Model} describes the proposed methodology, datasets, and neural network design. Section \ref{sec:Experiment} addresses the setup of the experiments conducted, while Section \ref{sec:Results} provides and discusses the experimental findings from the toxicity detection and compression methods used. Finally, conclusions and future work directions are noted in Section \ref{sec:Conclusion}. 


\section{Background and Related work} \label{sec:background}

Speech recognition and audio processing tasks have traditionally relied on Hidden Markov models (HMMs) and recurrent neural networks (RNNs) \cite{HMMRNN}. However, these models have limitations, such as difficulty in capturing long-term dependencies and limited scalability \cite{limitationrnn}. Recently, transformers have emerged as a promising alternative for speech recognition and audio processing tasks \cite{latif2023transformers}. They have demonstrated impressive results, including identifying the musical genre of a track \cite{transformermusic}, detecting environmental sounds \cite{transformersenvsounds}, and recognizing emotional sentiment \cite{transformersemotionalrecognition}. Transformers' effectiveness can be attributed to their ability to capture contextual information and long-term dependencies in sequence data \cite{transformers2020}. The key innovation of the transformer architecture is the attention mechanism, which allows the network to selectively focus on different parts of the input sequence when computing the output. Unlike traditional sequence-to-sequence models that rely on RNNs to process the input sequentially, transformers can process the entire sequence in parallel, leading to state-of-the-art performance in many audio processing and speech recognition tasks.

In the remainder of this section, we provide background and related work for wav2vec2.0 (\S \ref{wave2vec}), multi-task learning (\S \ref{MTLBackground}), toxicity detection (\S \ref{toxicityDetection}), and transformers for edge computing (\S \ref{tinyML}).

\subsection{Wav2vec2.0}   \label{wave2vec}



Wav2vec2.0 is a transformer-based model pre-trained in a self-supervised approach to learn representations directly from raw audio signals \cite{wav2vec2020}. The model is composed of multiple layers of CNNs for local feature extraction and a transformer-based architecture for global context modeling. During pre-training, the model is trained on a large corpus of unlabeled audio data (960 hours from the LibriSpeech dataset) to predict masked portions of the audio waveform using a contrastive loss function, similar to the pre-training phase in NLP where a model is trained to predict masked words in a sentence. This approach encourages the model to generate similar vector representations for semantically related frames and dissimilar representations for unrelated frames. 
Two versions of the Wav2vec2.0 model have been released for public use, a base model and a large one. The base model has 12 transformer layers with eight attention heads each and 768 hidden units, while the large model has 24 transformer layers with 16 attention heads each and 1024 hidden units. The wav2vec2.0 model has attracted significant interest due to its ability to achieve state-of-the-art results on various speech-related tasks while using considerably less labeled data than previous approaches. The success of the model is attributed to its pre-training phase, which generates a vector representation of the audio signal capturing critical speech-related features, such as phonemes. This representation can be fine-tuned using labeled data for specific downstream tasks, such as speech recognition \cite{wav2vecASR}, emotion detection \cite{wav2vecemotion}, and speaker recognition \cite{wav2vecspeaker}, resulting in improved accuracy on these tasks. We use Wav2vec2.0 as the foundation model in our proposed model.

\subsection{Multi-task learning}
\label{MTLBackground}

To improve the performance of downstream tasks, researchers have explored additional techniques such as Multi-task learning (MTL) which involves training a model on multiple related tasks simultaneously to leverage the commonalities between them and improve overall performance \cite{multitask1}. It has been successfully applied in many areas of ML, including NLP \cite{multitasknlp}, computer vision \cite{multitaskcomputervision}, and speech recognition \cite{multitaskspeech}.
 Typically, one main task and one or more auxiliary tasks are learned. The auxiliary tasks help the model to learn better and converge faster, resulting in better performance for the main task. In previous work, Ruder et al. \cite{mutitask-relatedwork} explored the effectiveness of multitasking learning for NLP tasks and observed that the joint training of related tasks, compared to separate training for different tasks, improved performance. However, they also noted that task dissimilarity and differences in data distribution could lead to negative transfer, reducing performance.
Another study by Alharbi and Lee \cite{multitask-relatedwork2} used MTL to improve emotional sentiment and sarcasm detection performance. The authors trained a single model to simultaneously predict the sentiment and sarcasm of text, leveraging the shared semantic information between the two tasks. Their results showed that MTL improved performance on both tasks compared to training on each task individually.

 


\subsection {Toxicity detection}
\label{toxicityDetection}

Toxic language is an umbrella term for various negative behaviors such as swearing, bullying, hate speech, harassment, etc. \cite{toxic-def}. In the existing research literature, many studies have attempted to address the problem of toxic language using various ML techniques. However, most of these systems have focused on analyzing text content or text embedded in images to detect toxic comments on online platforms \cite{textembimages}. Many of these studies have used traditional ML methods like Support Vector Machines (SVM), Naive Bayesian (NB) \cite{svm}, and some deep learning methods like CNN and RNN \cite{transformerwhy}. However, transformers-based methods have recently shown superior results due to their ability to capture long dependencies from the text data, making it a powerful tool for identifying toxic speech \cite{transformeroutperformer}. 

In response to the proliferation of media formats other than text for online consumption (e.g., audio), a few studies on audio-based methods have been conducted to identify toxic behavior \cite{baseline1audio} \cite{baseline2detoxy}. Due to the subjectivity of toxicity detection, the context of the speech is crucial in determining whether the content may be classified as toxic. Thus, the method's success relies heavily on the quality of the feature representation obtained. In the study \cite{baseline1audio} that aims to detect toxicity in online gaming, an audio-based toxic language classifier was proposed that used a self-attention mechanism to capture contextual representations of audio data. The authors compared two attention approaches and tested the proposed models on a private toxic-based dataset and a public emotional dataset called IEMOCAP. 
Their proposed method consists of two main stages. In the first stage, high-level feature extraction is performed through a CNN architecture; while in the second stage, these features are compressed through an attention mechanism. Two alternative attention techniques are investigated: (1) Learnable Query Attention and (2) Self-Attention. The results showed that Self Attention improves classification performance on their private corpus compared to the baseline.

In another important work \cite{baseline2detoxy}, Ghosh et al. proposed DeToxy, the first annotated dataset for toxic speech detection in the English language curated using 2 million utterances from a range of publicly available speech datasets integrating text and spoken cues. The authors compared several baseline methods for handling toxicity detection utilizing audio signals. The first approach is a 2-step method that uses the wav2vec2.0 model as an ASR system to generate transcripts from spoken utterances, followed by a sequence classification model based on a BERT base transformer architecture to categorize the potential toxicity of the transcript. The second approach is an E2E method that extracts log-Mel filter banks from the audio input sequence and calculates a global average pooling over the hidden states to obtain an embedding for each utterance. This embedding is then fed to the decoder to make the toxicity classification. The final strategy, which also falls under the E2E approach, involves employing the wav2vec2.0 model as a toxicity classifier. They achieved improved results using wav2vec2.0 model.
However, these methods suffer from two major flaws. Firstly, the complexity involved with utilizing two large models for toxicity detection. Secondly, the authors used an ASR system to generate text from audio input in a 2-step process, which can significantly reduce the system's performance in noisy environments \cite{ASRdegrades}.






\subsection{Transformers For Edge Computing}
\label{tinyML}
Despite transformer-based models achieving state-of-the-art performance in various domains such as vision, NLP, and speech recognition, their computational complexity due to the vast number of parameters, in the order of billions, makes them unsuitable for low-capability devices or latency-sensitive applications. To address this issue, researchers have been exploring various approaches to develop lightweight transformer models that can achieve high performance with fewer parameters and lower computation requirements. 

Several techniques have been proposed for compressing and optimizing these models without significantly sacrificing accuracy. Pruning, quantification, and, more recently, knowledge distillation are the most commonly used techniques \cite{compressionmethods1} \cite{compressionmethods2}. Pruning involves removing unimportant weights or connections in a trained model. Quantization reduces the precision of numerical values in a model, typically from 32-bit floating-point to lower-precision fixed-point or integer values, to reduce memory usage and increase inference speed. Knowledge distillation trains a smaller model (the ``student") to mimic the behavior of a larger, more accurate model (the ``teacher") by using the teacher's output probabilities as soft targets during training. This technique can create a more efficient model with a smaller memory footprint. 
These techniques provide opportunities for lightweight models to be deployed on low-power devices without relying on cloud-based processing, which is especially crucial for applications in smart homes, smart cars, agriculture, and healthcare \cite{warden2019tinyml}. 

\begin{figure*}[!t]
  \centering
  \includegraphics[width=0.9\linewidth]{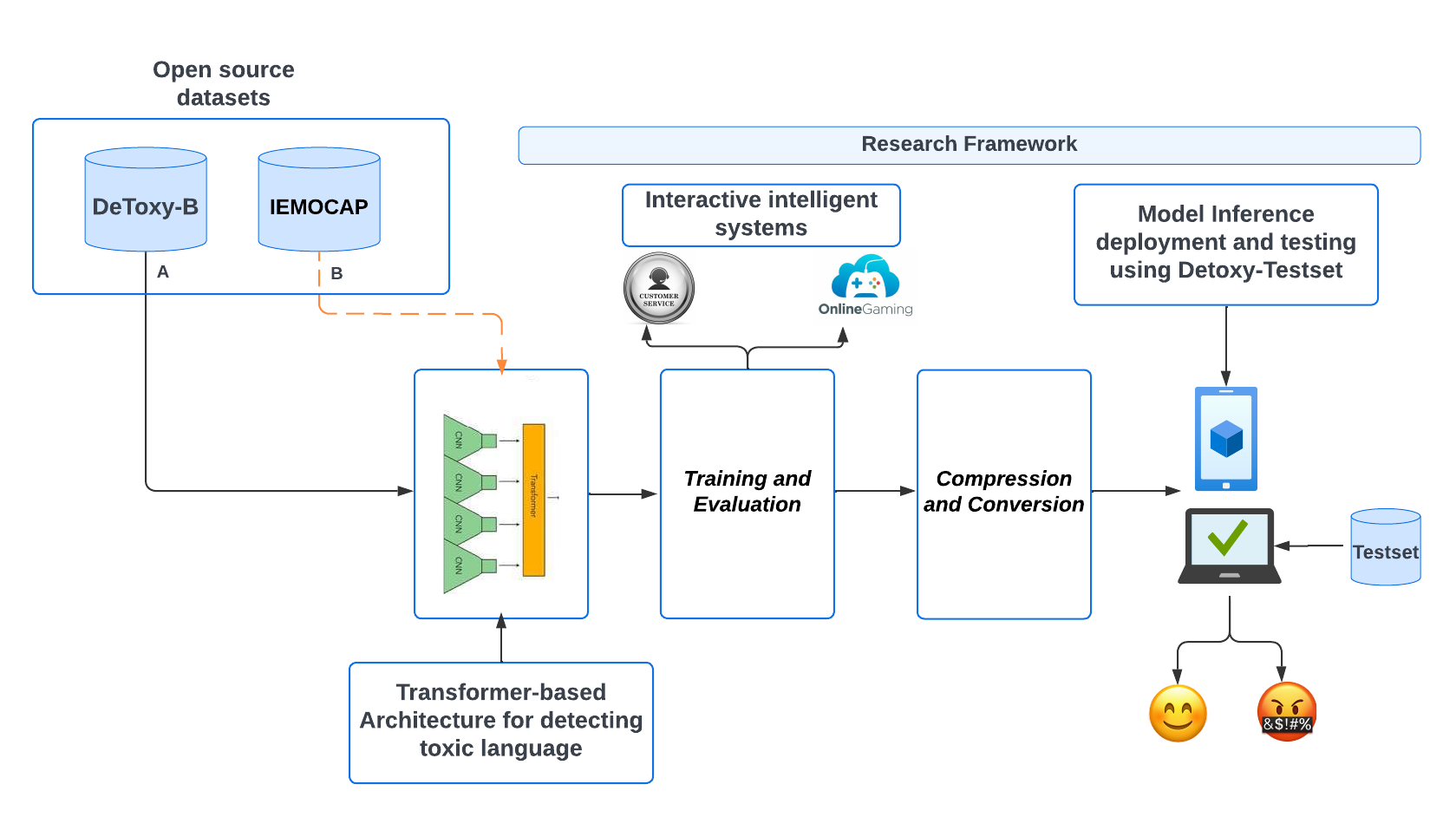}
  \caption{The high-level design of the proposed framework, where dataset A is the DeToxy dataset used to train and evaluate the model, and B is another dataset used to demonstrate the effectiveness of the suggested model. After being trained and evaluated, the model undergoes conversion and compression to make it efficient enough to be used on an edge device.}
  \label{framework}
\end{figure*}

In \cite{bertcompressing}, authors investigate the efficacy of multiple compression techniques, including pruning, quantization, and knowledge distillation, on the BERT model concerning NLP tasks like sentence pair classification, single sentence classification, and machine reading comprehension. The study assesses the performance of the compressed models on two benchmarks, GLUE and SQuAD, and includes an analysis of their speed on both GPU and CPU devices. 
The authors in \cite{Mobileformer} propose Mobile-Former architecture that merges MobileNet and Transformer using a two-way bridge, combining local and global features. The Transformer element of Mobile-Former is lightweight, with only a small number of tokens that are randomly initialized to learn global priors, making it computationally efficient. 
Another research proposed W2V2-Light, a lightweight version of wav2vec 2.0, to overcome hardware memory limitations \cite{lightwav2vec}. The authors conducted evaluations of their approach for speech recognition tasks using the standard LibriSpeech dataset. They introduced two sharing methods, namely Layer-wise Parameter Sharing (LPS) and Attention Alignment Sharing (AAS), to reduce memory consumption and computational costs. LPS involves sharing the same parameters across all Transformer layers to reduce the model size, while AAS pre-computes the attention alignment in the first Transformer layer and reuses it across all layers to mitigate the quadratic time complexity of softmax attention. W2V2-Light achieved a model size of 28M and a 1.31x speedup compared to wav2vec2.0. In contrast, our study utilized quantization and knowledge distillation techniques to compress wav2vec2.0 for toxicity classification tasks, resulting in a smaller model size of 25M and a 2x speedup. 

\textit{In this work, we utilize a single wav2vec2.0 model trained on both toxicity and ASR tasks simultaneously using the MLT technique to improve the contextual feature extraction ability of the model and compress the model to restrain the model complexity so that they become deployable on edge devices.}

\section{Proposed Framework} \label{sec:Proposed-Model}

In this section, we present the main components of the proposed framework, as depicted in Figure \ref{framework}. The sequential pipeline is designed to exploit the advantages of the pre-trained wav2vec2.0 model, multitasking technique, and compression techniques, enabling the efficient development of a toxicity detection model that can achieve state-of-the-art performance in both online and real-world settings.

\subsection{Proposed Multitask Model } \label{sec-proposed-arch}

\begin{figure}[!t]
  \centering
  \includegraphics[width=1\linewidth]{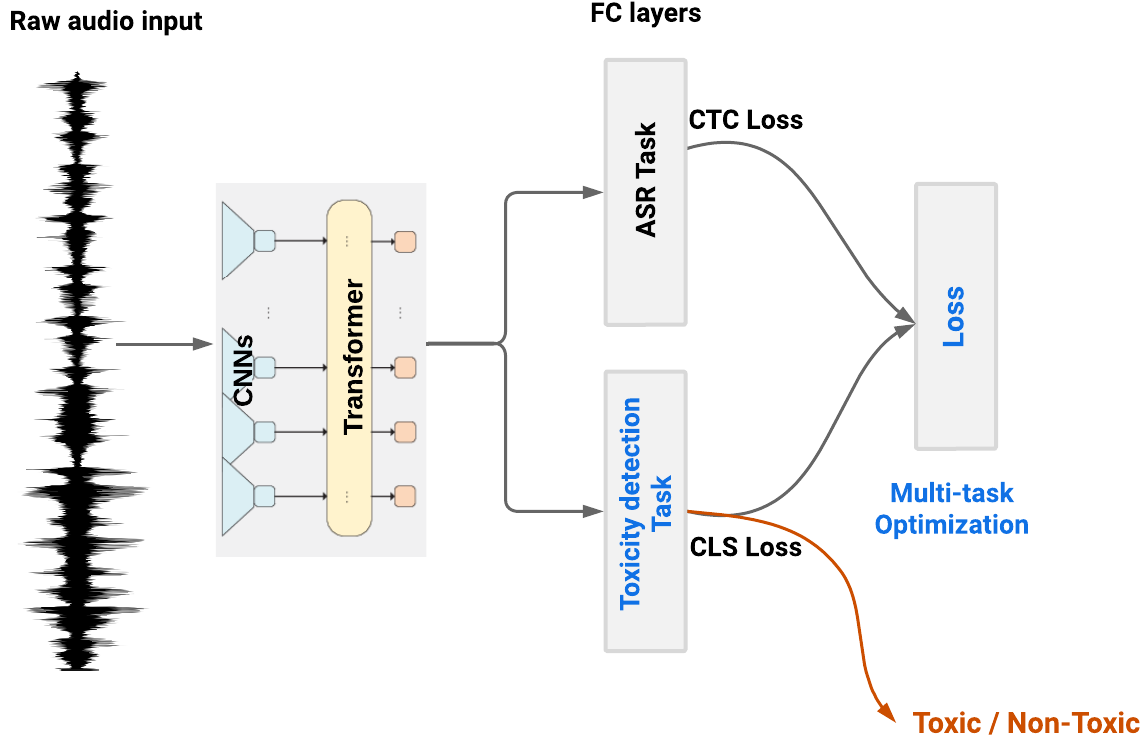}
  \caption{The proposed architecture used to detect toxicity in audio waves and an illustration of multi-task learning used to improve the wav2vec2.0 model's ability to capture contextual features. }
  \label{Model}
\end{figure}

We propose a multitask network that uses wav2vec2.0 as a shared encoder, as shown in Figure \ref{Model}. Wav2vec2.0 extracts high-level features using CNN layers from the raw speech, which are then fed into the transformer part to extract contextual embeddings. These embeddings are passed to two heads including ASR and toxicity classifier. The primary objective of the wav2vec2.0 study was to use the learned representations to boost ASR performance while using less training data via self-supervised learning. However, in this paper, we focus on improving the wav2vec2.0 model's capacity as a classifier by capturing generalised features in natural speech.
We added a fully connected layer to the wav2vec2.0 model and used the cross-entropy loss function to create a toxicity classifier. The output from wav2vec2.0 is first pooled using average pooling to get the 768-dimensional embedding, which is then projected into a 2-dimensional vector for classification as toxic or non-toxic.

We simultaneously trained the proposed model in multitask setting, with toxicity detection as the primary task and ASR as the auxiliary task. This helps enhance the performance of the toxicity detection model by capturing generalized contextual features.
We calculate the model loss by combining the classification (CLS) loss and the Connectionist Temporal Classification (CTC) loss. During training, the network's parameters are updated using backpropagation. Each task contributes to the cost function with a weighted term, which is determined by the following general equation:
\begin{equation}
\varepsilon_{MTL}= \varepsilon_{\text{Main}}+ \sum_{n=1}^{N}\lambda_{n}*\varepsilon_{\text{Auxiliary}_{n}},
\end{equation}
where the weight $\lambda_{n}$ represents the importance of the nth auxiliary task relative to the main task. A larger value of $ \lambda_{n} $ means that the auxiliary task has a greater impact on training, whereas a smaller value means that it has little effect. In our study, the equation will be as follows:

\begin{equation}
\label{mtl_eq}
\varepsilon_{\text{ToxicityDetection}}= \varepsilon_{\text{\text{CLS}}}+ (\lambda*\varepsilon_{\text{\text{ASR}}})
\end{equation}

At the inference time, only the network's outputs for the main task are used. Figure \ref{Model} shows the toxicity detection proposed model leveraging the MTL technique.

\subsection{Model Optimisation for Edge Devices} \label{sec-optimisation}



In order to optimize the model for edge devices, we utilize three compression techniques: quantization, pruning, and knowledge distillation. Quantization involves storing model weights in low-precision formats, which can accelerate operations on hardware with reduced precision support and reduce overall memory usage. Pruning reduces memory usage by setting neural network weights to zero, thereby removing unnecessary operations. Finally,
 knowledge distillation aims to obtain a simple, lightweight model while maintaining its accuracy by extracting knowledge from a large, complex teacher to a small, simple student model. As highlighted in \cite{wagner2023dawn}, larger architectures do not always lead to better performance. These models have higher knowledge capacity that capacity might not be fully utilized. Therefore, knowledge distillation can help transfer knowledge from a large model to a smaller one. 

For knowledge distillation, we utilise our proposed MTL-wav2vec2.0 model to serve as the teacher model with 12 transformer layers since we use the wav2vec2.0 base version. On the other hand, the student model uses the same wav2vec2.0 model as the teacher, but with five layers of transformer component and the same number of CNN layers. The number of CNN or transformer layers can be reduced when constructing the student model. Since there are only seven CNN layers, and they are not a significant computational bottleneck in comparison to transformer layers, we chose not to reduce them. Through knowledge distillation training, information is condensed and transferred from the big teacher model to the small student model.

\begin{figure}[t]
  \centering
  \includegraphics[width=1\linewidth]{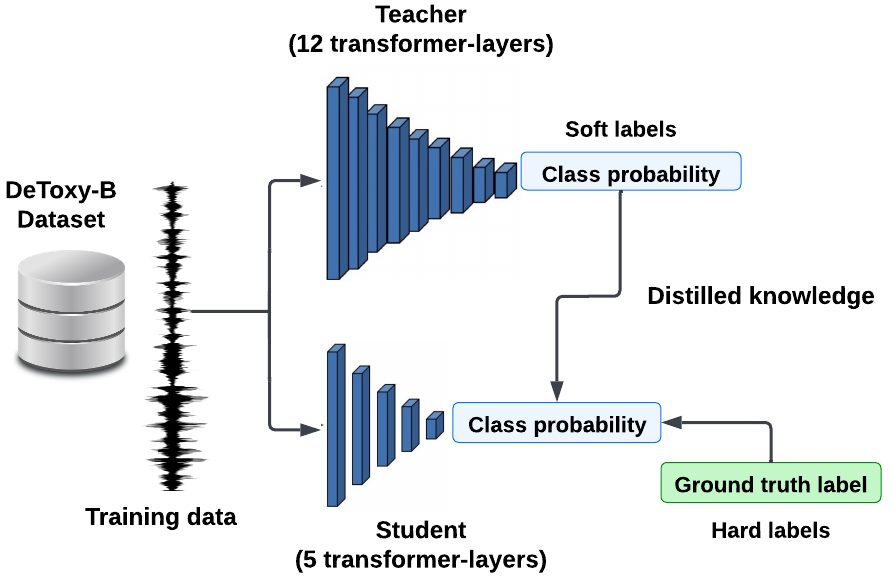}
  \caption{Knowledge distillation training process applied on the MTL wav2vec2.0 model to obtain a lightweight small model}
  \label{Knowledge-D}
\end{figure}

Figure \ref{Knowledge-D} depicts the training process for knowledge distillation. The teacher and student models receive the same input data, and the loss is then calculated using the results of both models. We perform gradient descent on the determined loss to update the student model. The teacher model is not updated by not calculating the gradient for its loss. The final knowledge distillation loss is the difference in probability distributions between the teacher and student models.

\section{Experiments} \label{sec:Experiment}


\label{sec-f-experiments}

\subsection{Datasets} \label{sec-dataset}

\subsubsection{DeToxy-B}
The primary dataset on which our proposed model is evaluated is DeToxy-B \cite{baseline2detoxy}. This dataset is a newer version of another dataset called the DeToxy dataset. DeToxy is the only publicly available dataset designed for speech toxicity classification in conversational utterances in the English language. It is a multimodal dataset that comprises speech and text cues. Three professional annotators manually annotated the audio clips as toxic or non-toxic. Then, a simple majority voting among the three annotations was used to determine the final toxicity class of each utterance. The audio clips were compiled from various open-source datasets; Table \ref{DeToxy-table} contains its description of the statistical data. By incorporating data from different datasets, DeToxy has diverse types of data from multiple contexts and sources, such as conversational and question-answering data. Consequently, the DeToxy dataset can be effectively applied in real-world contexts beyond its original intended use in online settings. Additionally, The DeToxy dataset includes speakers of various genders, ages, accents, and ethnicities. The authors have created a new version of the DeToxy dataset called DeToxy-B, which is designed to be more balanced than the original dataset used in this study. 

\begin{table}[tbp]
\centering
\scriptsize
\caption{DeToxy-B statistics }
\label{DeToxy-table}
\begin{tabular}{lcccc}
\hline
\textbf{Dataset Name} & \textbf{\ Audio Clips} & \textbf{\ Toxic} & \textbf{\ Non-Toxic} & \textbf{Time Duration} \\ \hline
\textbf{Common Voice \cite{commonvoice}} & 11,551 & 2,888 & 8,663 & 12: 38: 17 \\
\textbf{IEMOCAP \cite{iemocap}} & 1,090 & 274 & 816 & 01: 19: 26 \\
\textbf{LJ Speech \cite{LJSpeech}} & 148 & 40 & 108 & 00: 14:57 \\
\textbf{MELD \cite{meld}} & 565 & 142 & 423 & 00: 31: 05 \\
\textbf{MSP-Improv \cite{MSPImprov}} & 523 & 129 & 394 & 00: 36: 32 \\
\textbf{MSP-Podcast \cite{MSPPodcast}} & 2,772 & 692 & 2,080 & 04: 01: 57 \\
\textbf{Social-IQ \cite{socialIQ}} & 479 & 122 & 357 & 00: 36: 40 \\
\textbf{Switch board \cite{switchboard}} & 1,824 & 456 & 1,368 & 02: 28: 57 \\
\textbf{VCTK \cite{VCTK}} & 199 & 50 & 149 & 00: 08: 51 \\ \hline
\textbf{Total} & 19,151 & 4793 & 14,358 & 22: 36: 42 \\ \hline
\end{tabular}
\end{table}

\subsubsection{IEMOCAP} 
We further evaluated our proposed model on another public dataset, IEMOCAP \cite{iemocap}, primarily annotated for emotional sentiment detection, to better illustrate the suggested model's effectiveness. 
We employed the same methodology adopted by Midia and Dimitra in their research \cite{baseline1audio}. They changed the labels of the audio clips to more closely resemble the toxicity detection problem, where happy and excited emotion groups were combined into an un-toxic class. In contrast, frustrated and angry emotion groups were combined into a Toxic class. 

\subsection{Training Protocol}



In the \textit{first experiment}, we train our model without the ASR as an auxiliary task, we set the weight coefficient $ \lambda$ to 0 (equation \ref{mtl_eq}), indicating that the auxiliary task had no impact on the training process. Consequently, we replicated the DeToxy-B paper's method by training the model in a Single Task Learning manner to classify the utterances as toxic and non-toxic. 

In the \textit{second experiment}, we train wav2vec 2.0 using the MTL technique since we have the ground truth label for both tasks. The main task is toxicity classification, and the auxiliary task is the ASR task. The weight coefficient $ \lambda $ is set to 0.1 to test the impact of the ASR task as an auxiliary task on the model's performance on the toxicity classification task. We ran grid search experiments to determine the ideal value for $ \lambda$, as in \cite{multitaskspeech}, and we found that $ \lambda $ = 0.1 performs the best.  

In the \textit{third experiment}, we establish an additional baseline to further evaluate our proposed MTL model  using the IEMOCAP dataset. To build and evaluate our models, we use the PyTorch 1.7 Framework. The pre-trained wav2vec2.0 implementations and tokenizers are obtained from the HuggingFace library \cite{wav2vecBASE}. We train all our models with Adam optimizer in batched mode with a batch size of 2 and a learning rate of 5e-5 for 100 epochs. 
The configuration of the environment in which we trained and evaluated the models and conducted experiments is described in Table \ref{Table5-1}

In the \textit{fourth experiment}, we use the Neural Network Exchange (ONNX) library\ref{https://onnx.ai/} to convert our trained model to the ONNX standard format and then employ quantization and pruning. We quantize the model into in8 and perform the pruning with a sparsity level of 50\%. We run an inference of this model on a CPU-based laptop; its specifications are listed in Table \ref{Table5-2}. 

Two crucial parameters must be set to apply knowledge distillation: \textit{alpha} ($\alpha$) and \textit{temperature} ($T$). These factors are crucial for successful knowledge transfer from the teacher to the student model. The $\alpha$ parameter scales how much knowledge we should derive from soft and hard labeling. In many circumstances, the probability distribution of the teacher softmax has the correct class at a very high probability, with the other class probabilities very close to zero, which does not provide much information beyond the already provided ground truth labels. It is where the second parameter, $T$ comes into play. The softness of the teacher distribution can be improved by raising the $T$ in the teacher's objective function. Hence, when $T$=1, we obtain the typical softmax function. However, as $T$ increases, the probability distribution produced by the softmax softens, giving us additional details about which classes the teacher thought were more resembling the predicted class. This is referred to as the dark knowledge embedded inside the teacher model and transferred to the student model throughout the distillation process.

\begin{table}[!t]
\centering
\caption{Settings for Toxicity Detection Experiments}
\label{Table5-1}
\begin{tabular}{l|l}
\hline
\textbf{Operating System} & Ubuntu18.04- Linux \\ \hline
\textbf{GPU} & RTX3090 \\
\textbf{Number of CPU} & 24 \\
\textbf{Framework} & Python 3.9- PyTorch 1.7 \\ \hline
\end{tabular}
\end{table}
In our experiment, we set $\alpha$ to 0.5 and $T$ to 4.0. After applying the compressing methods on our MTL wav2vec2.0 model and getting a lightweight model, the DeToxy-B test set and CPU-based laptop are used for evaluation. Table \ref{Table5-2} lists the relevant device specifications.

\begin{table}[htbp]
\centering
\caption{Settings for Inference Experiments}
\label{Table5-2}
\begin{tabular}{l|l}
\hline
\textbf{Operating System} & MacOS Ventura 13.0 \\ \hline
\textbf{CPU} & 10 Core CPU- Arm Processor \\
\textbf{Framework} & Python 3.9\\ \hline
\end{tabular}
\end{table}

To investigate the toxicity classification inference, we configured the Python script to run on CPU cores. We analyzed the associated tradeoffs in terms of the F1 score, prediction time (latency), compressed model size, and RAM use.

\section{Results and Discussion}   \label{sec:Results}
We evaluate the different models and compared the performance using the accuracy, macro Average F1-score, precision, and recall metrics. In addition, we evaluate the performance of the models after compression by considering the inference runtime, model size, and memory usage. We initiate our experiments by investigating the performance of the wav2vec2.0 model on the DeToxy-B dataset, with the goal of establishing a baseline for subsequent experiments. The evaluation aimed to replicate the results reported in the baseline \cite{baseline2detoxy}. To ensure parity with the published results, we employ the same training, validation, and testing datasets created by the authors. The DeToxy-B dataset was divided into three sets, with 70\% of the randomly selected data used for training and validation, while 15\% was reserved for testing and evaluation purposes. It is worth noting that the DeToxy-B dataset included two datasets that we could not integrate into our study due to the unavailability of the raw data. Therefore, we replicate the baseline using the available datasets to ensure consistency with the DeToxy-B paper.

\subsection{Wav2vec2.0 model for Speech Toxicity Detection}

Since the transformer-based architecture of wav2vec2.0 is advantageous for capturing long-range dependencies in audio signals, which is essential for detecting the context and meaning of spoken language. The inclusion of CNN layers further enhances the model's ability to capture local features and patterns in the audio signal. This combination of CNN and transformer layers makes wav2vec well-suited for detecting the nuances of speech, such as tonal, emotional, and contextual features, which are critical for identifying toxic speech. Therefore, the performance of the wav2vec2.0 model in detecting toxicity using audio data on the DeToxy-B test set is evaluated in the first experiment, replicating the first baseline. As the findings presented in Table \ref{Table6-1}, our model achieved a relatively high average F1-score, which closely resembles the value stated in the paper (86.9 F1-score). This indicates that the model is effective in detecting the presence of toxicity in audio data. 

Nevertheless, considering that the wav2vec2.0 model is pre-trained to predict masked speech representations based on the surrounding context, which may be a specific sound or phoneme. We can improve its ability to learn more contextual features by learning to map the input speech signal to its corresponding text transcription via the ASR task. However, we must ensure that our modifications do not add unnecessary complexity to the model. Therefore, in our second experiment, we employed the MTL strategy with the ASR task serving as an auxiliary task while the main task is toxicity detection. We can deduce from the results in Table \ref{Table6-1} that the wav2vec2.0 model's ability to extract more contextual features increases as the results reflect where the MTL wav2vec2.0 performed 3.8 \% better than the plain wav2vec2.0 model.


\begin{table}[htbp]
\centering
\caption{Performance of the wav2vec2.0 model on detecting toxicity using the DeToxy-B dataset before and after applying the MTL technique. \textit{MTL-wav2vec2.0 model has a higher level of accuracy, indicating that MTL on both ASR and toxicity tasks enhances the model's ability to extract contextual features.} }
\label{Table6-1}

\begin{tabular}{lccrc}
\hline
\textbf{Model} & \textbf{Precision} & \textbf{Recall} & \textbf{Accuracy} & \textbf{Avg. F1-Score} \\ \hline
\textbf{STL-wav2vec2.0} & 85.9 \% & 87.2 \% & 89.7 \% & 86.5 \% \\
\textbf{MTL-wav2vec2.0} & 90.4 \% & 89.3 \% & 92.4 \% & 90.3 \% \\ \hline
\end{tabular}
\end{table}

\begin{figure} [htbp]
    \centering
  \subfloat[CNN features using PCA\label{CNN Features-PCA}]{%
       \includegraphics[width=0.5\linewidth]{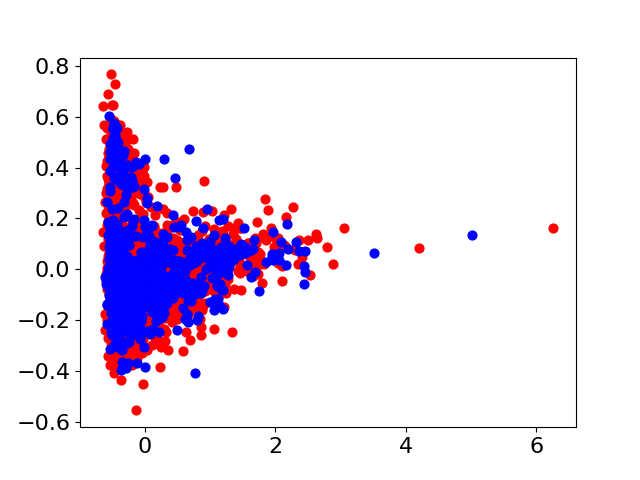}}
    \hfill
  \subfloat[Transformer features using PCA\label{Transformer features-PCA}]{%
        \includegraphics[width=0.5\linewidth]{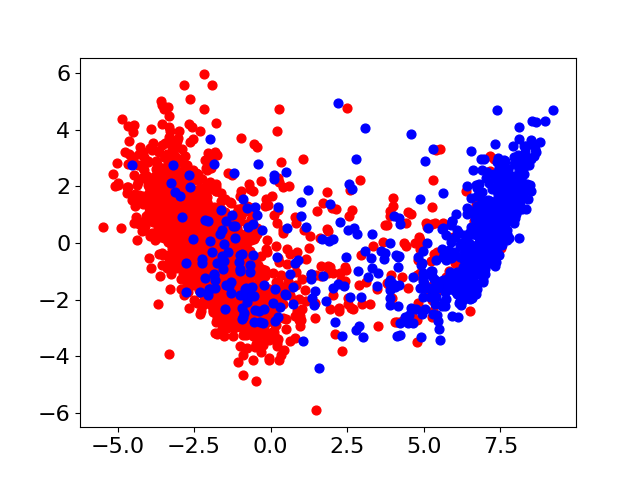}}
    \\
  \subfloat[CNN features using t-SNE\label{CNN features T-SNE}]{%
        \includegraphics[width=0.5\linewidth]{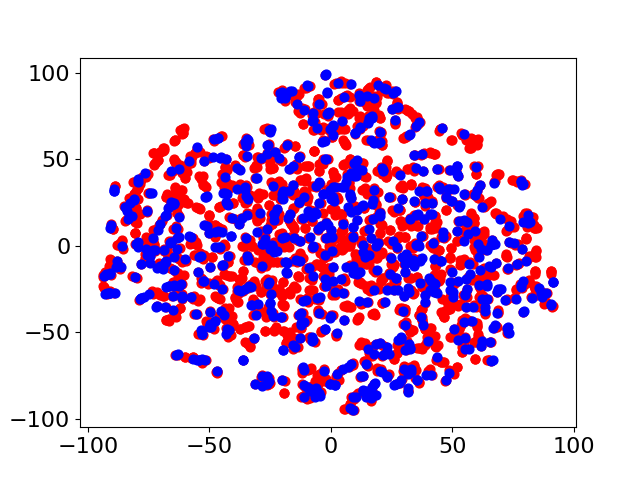}}
    \hfill
  \subfloat[Transformer features using t-SNE\label{Transformer features T-SNE}]{%
        \includegraphics[width=0.5\linewidth]{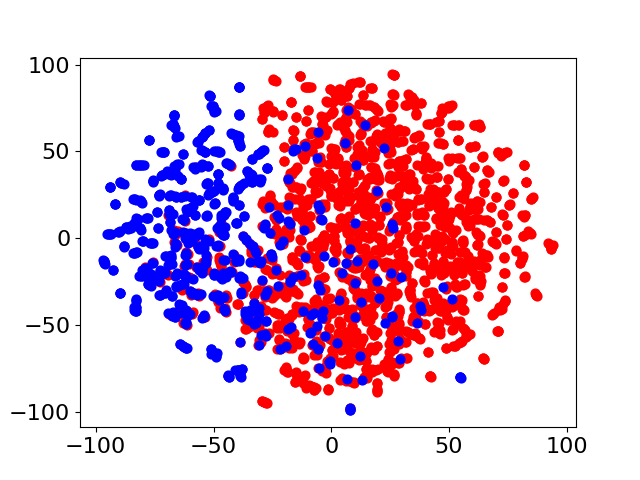}}
  \caption{Visualization of feature space separability for the DeToxy-B dataset.
        \textit{The two columns in the representation obtained from the MTL-wav2vec2.0 model correspond to the features extracted by its CNN and transformer parts, respectively. Clear separation of the blue and red classes indicates that the model effectively captures a high-quality representation of the audio samples.}}
  \label{visualization-4} 
\end{figure}

\begin{figure} [htbp]
    \centering
  \subfloat[STL\label{STL}]{%
        \includegraphics[width=0.5\linewidth]{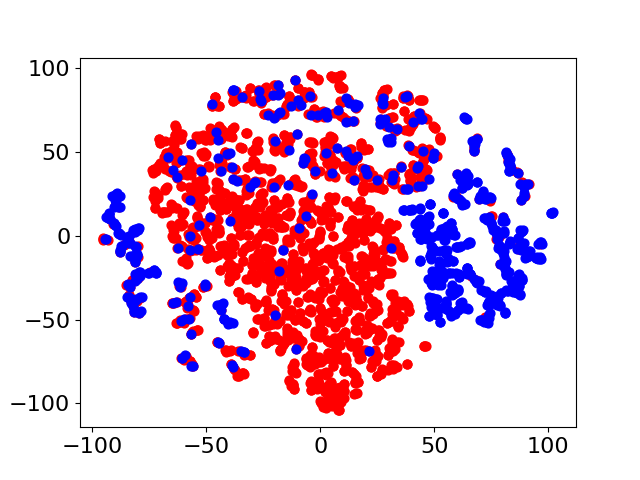}}
    \hfill
  \subfloat[MTL\label{MTL}]{%
        \includegraphics[width=0.5\linewidth]{Figures/Figure-T-SNE.png}}
  \caption{Visualization of feature space separability for the DeToxy-B dataset using the MTL and STL. \textit{The feature space comparison between MTL and STL models reveals that the MTL model's feature space is more separable, indicating its effectiveness at extracting relevant information for toxicity detection.}
 }
  \label{visualization-2} 
\end{figure}

In addition to the results presented in Table \ref{Table6-1}, we have visualized in Figure \ref{visualization-4} the feature vectors extracted from the CNN layers and the feature vectors extracted from the transformer layers by Principal Component Analysis (PCA) and t-distributed Stochastic Neighbor Embedding (t-SNE) reduction techniques, with red and blue colors representing the toxic and non-toxic classes. In both PCA and t-SNE graphs, the transformer derived higher-level features that are more divisible than the CNN features. Furthermore, we visualized the feature vectors extracted from the transformer layers of the STL wav2vec2.0 and MTL wav2vec2.0. As Figure \ref{visualization-2} shows, the feature space for the MTL model appears more separable, indicating a meaningful learned representation for the toxicity detection task.

\begin{figure}[!ht]
  \centering
  \includegraphics[width=1\linewidth]{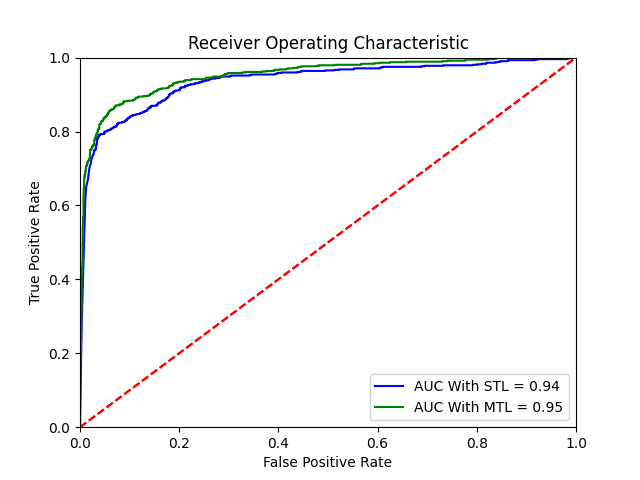}
  \caption{ROC of MTL and STL for DeToxy dataset. \textit{The higher AUC of the MTL-wav2vec2.0 model (compared to the STL-wav2vec2.0 model) provides evidence of its superior performance even in the presence of imbalanced datasets.}}
  \label{ROC}
\end{figure}

ROC curves are another method for comparing models in the situation of an imbalanced dataset. ROC is illustrated in Figure \ref{ROC}. Although the AUC in ROC for the MTL model is only 1\% higher than the STL model, this makes a significant difference in the greater accuracies, demonstrating MTL's ability to extract more relevant information.

\subsection{Comparison with State-of-the-Art Baseline on DeToxy}

The proposed model's performance is compared to other approaches by establishing two baselines, as described in Section \ref{sec-f-experiments}. To evaluate our results, we utilized the same metric that was used to evaluate each baseline in their respective studies. To establish the first baseline, we compared our proposed model MTL-Wav2vec2.0 to the model that yielded the best results in \cite{baseline2detoxy} on the DeToxy dataset. The researchers' top-performing model used a strategy similar to the STL-wav2vec2.0 but with embeddings from the ninth layer of the transformer component instead of the last layer of the wav2vec2.0.The comparison results presented in Table \ref{Table6-2} demonstrate that our proposed MTL model outperformed the best-performing model on the DeToxy dataset. Thus, concurrently training the model on two related tasks can improve its performance without introducing additional complexity.

\begin{table}[htbp]
\centering
\caption{Comparison between the baseline and our proposed model using the DeToxy-B dataset. \textit{MTL-wav2vec2.0 outperformed the first baseline, indicating better capture of relevant contextual features for toxicity problems.}}
\label{Table6-2}
\begin{tabular} {lc}
 \hline
\textbf{Model} & \textbf{Average F1-Score} \\  \hline
\textbf{MTL-wav2vec2.0} & 90.3 \% \\
\textbf{DeToxy-Baseline \cite{baseline2detoxy}} & 87.7 \% \\  \hline
\end{tabular}
\end{table}


\subsection{Comparison with State-of-the-Art Baseline on IEMOCAP}

In the third experiment, we established the second baseline, in which we make a direct comparison with the Self-Attentive CNN Hybrid model proposed in this work \cite{baseline1audio}. Following the paper's lead, we modify the IEMOCAP dataset's labels to align them more closely with the toxicity detection task and then use the modified dataset to train our MTL-wav2vec2.0 model. As the results show in Table \ref{Table6-3}, our model significantly outperformed their model. which is reasonable considering that their method only used one layer of self-attention, as compared to the 12 layers of transformers used in our model, which can extract a great deal of contextual, tonal, and emotional cues and are therefore essential in the task of detecting toxicity. We conducted an additional experiment to assess the effectiveness of the MTL approach by training the STL-wav2vec2.0 model on the IEMOCAP dataset. The results presented in Table \ref{Table6-3} indicate that our MTL-wav2vec2.0 model can extract more relevant features. It is worth noting that the results presented in Table \ref{Table6-3} of our baseline \cite{baseline1audio} were obtained after data augmentation, as the initial results were comparatively lower. In contrast, our model showed robustness to variations in audio signals without requiring additional data augmentation due to the self-supervised learning, where the wav2vec2.0 model is pre-trained on a hugely diverse range of audio data.

\begin{table*}[htbp]
\centering
\caption{Comparison between the second baseline ``Self-Attentive CNN Hybrid model'' and our proposed model ``MTL-wav2vec2.0'' using IEMOCAP dataset. \textit{The MTL-wav2vec2.0 model outperformed the second baseline model by a significant margin and showed better performance than the STL-wav2vec2.0 model on the second dataset.}}
\label{Table6-3}
\begin{tabular}{lclrc}
\hline
\textbf{Model} & \textbf{Precision} & \textbf{Recall} & \textbf{Accuracy} & \textbf{Weighted-Accuracy} \\ \hline
\textbf{STL-wav2vec2.0} & 81.51 \% & 87.49 \% & 85.45 \% & 84.80 \% \\
\textbf{MTL-wav2vec2.0} & 84.23 \% & 90.16 \% & 88.15 \% & 88.10 \% \\
\textbf{Self-Attentive CNN}\cite{baseline1audio} & 63.79 \% & 73.74 \% & 68.85 \% & 68.79 \% \\ \hline
\end{tabular}
\end{table*}


\begin{table*}[htbp]
\centering
\caption{Performance of MTL-wav2vec2.0 with and without quantization and knowledge distillation techniques, evaluated both individually and in combination. \textit{The compressed model obtained through the use of a combination of  compression techniques is a promising development towards deploying the model on edge devices.} }
\label{Table6-4}
\begin{threeparttable} 
\begin{tabular}{l c c c c r}
\toprule 
\hline
\textbf{Model}  & \textbf{Avg F1-score} & \textbf{Model Size} & \textbf{ Peak RAM} &\textbf{Inference Time} &\textbf{RTF}\\ \hline
\midrule 
\textbf{MTL-wav2vec2.0}  & 90.3 \% & 377.9 MB &  2.6 GB &  493 s & 0.0325\\ 
\textbf{Quan-MTLwav2vec}  & 89.4 \% & 95.2 MB &  0.8 GB & 1556 s & 0.057\\ 
\textbf{KD-MTLwav2vec}  & 81.9 \% & 102 MB & 1.4 GB &  300 s  & 0.022\\ 
\textbf{Quan-KD-MTLwav2vec}  & 78.1 \% & 25.9 MB & 0.6 GB & 956 s  & 0.04\\ \hline
\bottomrule 
\end{tabular}
\end{threeparttable}

\end{table*}


\subsection{Model Optimization} \label{sec-compressed}

In these experiments, we optimize our trained model using quantization, pruning, knowledge distillation, and their combination. For the evaluation, a laptop with a CPU will be used. This configuration choice is intended to simulate environments similar to those found in embedded devices, such as Raspberry Pi and mobile devices. To apply the quantization and pruning techniques, we used the Open Neural Network Exchange (ONNX) library for three key reasons: interoperability, hardware optimization accessibility \cite{ONNXportablity}, and compared to other Inference Engine (IE) libraries, it performed the best in terms of inference time and model footprint, especially for large models \cite{IEcomparison}. ONNX runtime supports many Deep Neural Network (DNN) model types and integrates with accelerators on a wide range of hardware. It is possible to convert models from various frameworks to the universal ONNX format. Once the models have been transformed into the ONNX format, they can be executed on many platforms and devices.


According to experimental findings, pruning the MTL-Wav2vec 2.0 model at a sparsity level of 50\% did not result in any improvement in inference speed or model size. This result was consistent with previous studies, which showed that pruning only eliminates a portion of the model's weights without reducing the overall computational complexity of the model \cite{pruning}. As presented in Table \ref{Table6-4}, the quantized model is approximately four times smaller than the original model and reduces peak RAM usage by 3.25, but it does not lead to faster inferences. Nevertheless, the quantized model demonstrates good performance, with only a 1\% decrease in the macro F1-Score. Therefore, the quantized model can be beneficial when model size is a concern.

\begin{figure}[!h]
    \centering

  \subfloat[MTL-wav2vec2.0\label{RAM-Original}]{%
        \includegraphics[width=1\linewidth]{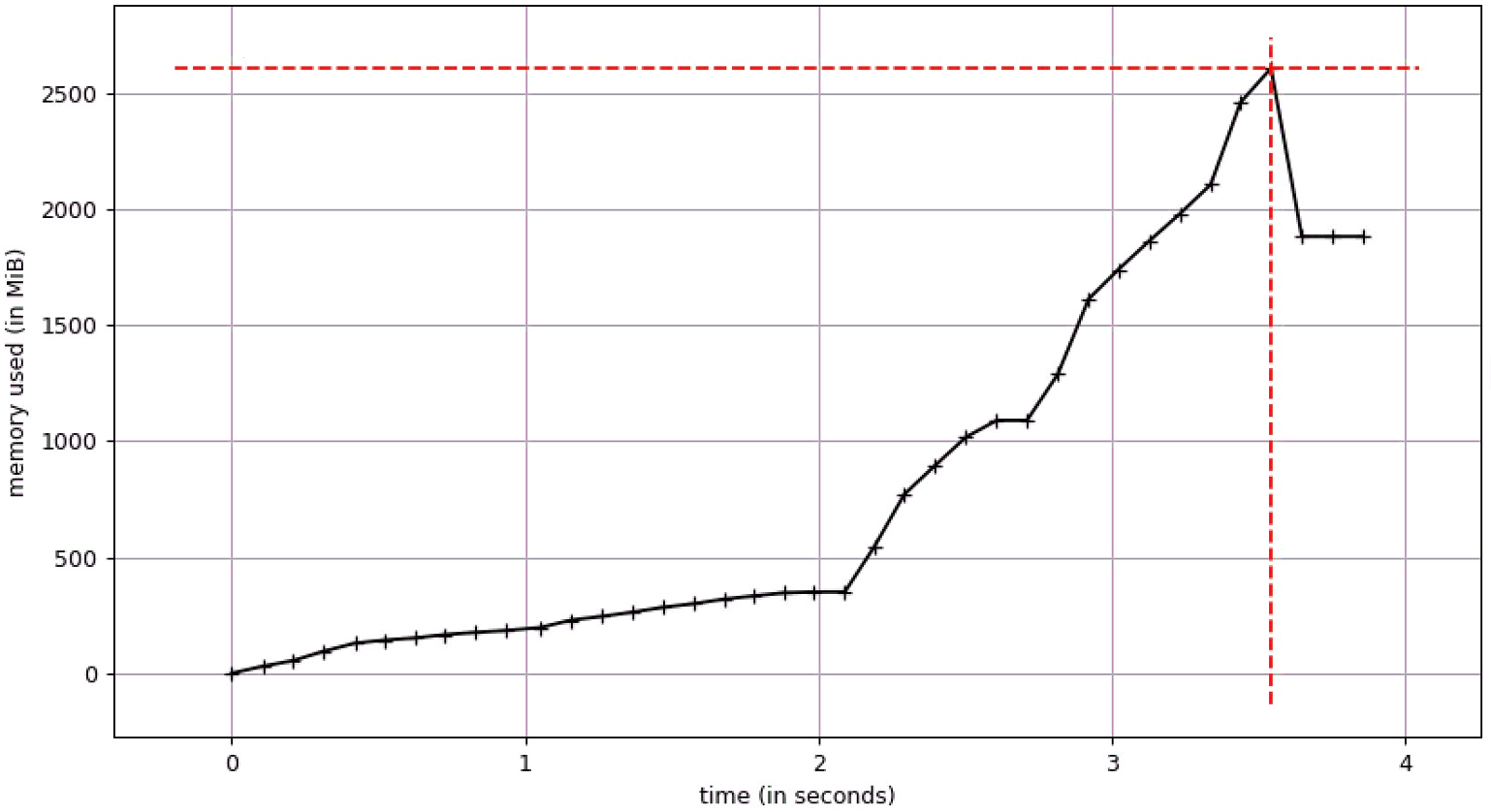}}
    
  \subfloat[Quan-KD-MTLwav2vec2.0\label{RAM-KDQUANT}]{%
        \includegraphics[width=1\linewidth]{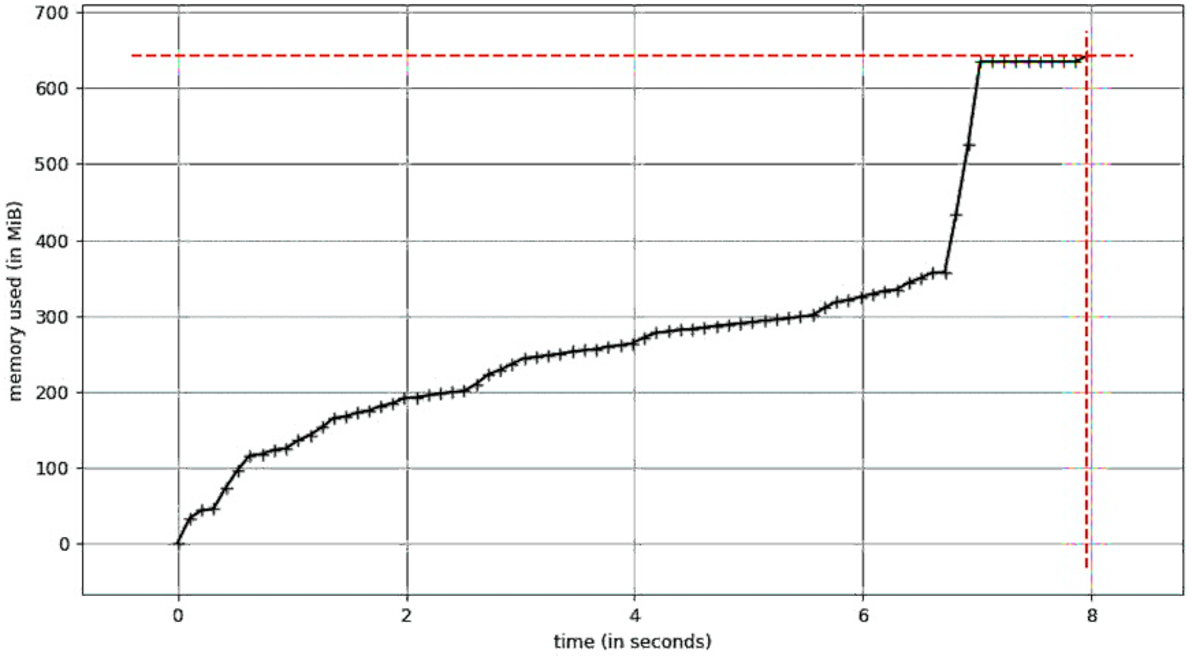}}
 \caption{The RAM used for the original model MTL-wav2vec2.0, and the Quan-KD-MTLwav2vec2.0 inference. \textit{The combination of knowledge distillation and quantization techniques has reduced the model's RAM usage by 4.3 folds compared to its original size.}} 
  \label{RAMUsage} 
\end{figure}

In contrast, knowledge distillation produces a compact model with fast inference but at a significant performance penalty. The student model, which is 3.7 times smaller and 1.6 times faster than the original model, resulted in an 8\% decrease in the average F1 score. Combining quantization and knowledge distillation techniques resulted in a model with a 14.6-fold smaller hard disk footprint than the original model. Additionally, the memory footprint is reduced by 4.3 times compared to the original model, as depicted in Figure \ref{RAMUsage}. However, it should be noted that this was accomplished at the expense of the model's accuracy.


It is important to note that the inference time is calculated for all samples in the test set. However, to obtain a more accurate measurement of the inference time taken by each audio input, particularly in the context of real-time processing, the Real-Time Factor (RTF) can be calculated. This is achieved by dividing the inference time by the duration of the input utterance while taking into account the varying durations of the samples in DeToxy-B. Using this method to calculate the RTF provides a more precise measure of the inference time taken for each input second. For instance, if an audio sample has a duration of 6 seconds, the original model's inference time will be 0.2 seconds, while the KD model's inference time will be 0.13 seconds.

Consequently, when conducting inference on a CPU-based laptop, the compressed model (Quan-KD-MTL wav2vec2) demonstrates a reduction in the hard disk footprint by a factor of 14.6. Moreover, it has a memory footprint of approximately 25 MB and a real-time factor of less than 1.0. These characteristics make the latest transformer models suitable for deployment on typical edge devices, enabling private, secure, reliable, and always-available toxicity detection processing.

However, the decision of which model to choose is influenced by the trade-off between the edge device's capabilities and the priority assigned to accuracy. If the device has ample processing power and memory, a quantized wav2vec2.0 model can be utilized to achieve high accuracy in toxicity detection. However, for devices with limited memory, a combined model that employs both quantization and knowledge distillation techniques may be preferred, even though this could result in a reduction in accuracy.

All models have a real-time factor (RTF) of less than 1, indicating that they are suitable for real-time processing. However, performance may be affected if the device has fewer CPU cores, such as in the case of some embedded devices. Nonetheless, these models have been demonstrated to work in real-time on CPU-based devices, such as minicomputers or Raspberry Pi.




\subsection{Limitations}

The study has some limitations that need to be addressed. Firstly, the proposed solution may not be suitable for streaming mode, which may result in insufficient contextual information for the model to accurately detect instances of toxic language. Secondly, deploying the proposed system in an ambient setting where multiple speakers are present poses unique challenges, particularly in identifying the specific speaker responsible for engaging in toxic speech. To address these limitations, future research may consider incorporating speaker recognition in overlapping regions to identify the individual responsible for the toxic utterance and collect a dataset from the same domain in which the system will be deployed. These extensions could potentially improve the accuracy and effectiveness of automated tools for detecting and addressing instances of toxic language in real-time while also accounting for the specific contextual factors of the deployment environment.

\subsection{Future Work}
To address the aforementioned limitations, future research may consider incorporating speaker recognition in overlapping regions to identify the individual responsible for the toxic utterance rather than solely providing a toxicity classification. Additionally, researchers may collect a dataset from the same domain in which the system will be deployed, such as offices, hospitals, or schools, as each environment has distinct characteristics and ambient noise levels that may impact performance. These extensions could potentially improve the accuracy and effectiveness of automated tools for detecting and addressing instances of toxic language in real-time while also accounting for the specific contextual factors of the deployment environment.

\section{Conclusions}  \label{sec:Conclusion}



This study presents a novel transformer-based methodology for detecting toxic language using only audio data. 
The proposed solution employs multitask learning, outperforming state-of-the-art methods with a 90.3\% macro F1 score on the DeToxy-B dataset. This solution can be used for interactive intelligent systems with a wide range of uses, including content moderation, online gaming, and customer service.
We optimise the proposed model for edge devices applied and used three compression techniques, pruning, quantization and knowledge distillation. We used CPU-based inferencing to evaluate the performance of our compressed model. The compression technique chosen is determined by the priorities required by the end application. Pruning did not cause a substantial change in the complexity of the original model. In contrast, Knowledge distillation produces a compressed model with fast inference but at the expense of some performance. The quantized model has no performance loss, but its inference speed is not faster. Additionally, we combined the two techniques to create an optimal balance between model size and inference time; however, this came at the expense of some decrease in the model's accuracy. Given the model's compact size of 25 MB, RAM consumption of 0.6 GB, and fast inference time of 0.02 seconds for a one-second input, the model's inference can be readily deployed on a variety of resource-constrained devices. 

\vspace{10mm}



\end{document}